\documentclass[journal]{IEEEtran}

\usepackage{times}
\usepackage{soul}
\usepackage{url}
\usepackage[hidelinks]{hyperref}
\usepackage[utf8]{inputenc}
\usepackage[small]{caption}
\usepackage{graphicx}
\usepackage{amsmath}
\usepackage{amssymb}
\usepackage{amsthm}
\usepackage{booktabs}
\usepackage{algorithm}
\usepackage{algorithmic}
\usepackage{enumitem}
\usepackage{caption}
\usepackage{subcaption}
\usepackage{soul}
\usepackage{color, xcolor}

\title{Decoupling Speaker-Independent Emotions for Voice Conversion Via Source-Filter Networks}

\author{Zhaojie Luo,~\IEEEmembership{Member,~IEEE,}
        Shoufeng Lin,~\IEEEmembership{Senior Member,~IEEE,}
        Rui Liu,~\IEEEmembership{Member,~IEEE,}
        Jun Baba,~\IEEEmembership{Member,~IEEE,}
        Yuichiro Yoshikawa,~\IEEEmembership{Member,~IEEE,}
        Ishiguro Hiroshi,~\IEEEmembership{Member,~IEEE,}
\thanks{Z. Luo is with Graduate School of Engineer Science, Osaka University, Japan. e-mail: luo@irl.sys.es.osaka-u.ac.jp}
\thanks{S. Lin is with School of Electrical Engineering, Computing and Mathematical Sciences, Curtin University. e-mail: shoufeng.lin@graduate.curtin.edu.au }
\thanks{R. Liu is with National University of Singapore and Singapore University of Technology and Design (SUTD), Singapore. e-mail: liurui\_imu@163.com }
\thanks{Jun Baba are with CyberAgent, Inc. Tokyo, Japan.}
\thanks{Yuichiro Yoshikawa and Ishiguro Hiroshi are with Graduate School of Engineer Science, Osaka University, Japan.}}

\begin{document}

\maketitle
\begin{abstract}
Emotional voice conversion (VC) aims to convert a neutral voice to an emotional (e.g. happy) one while retaining the linguistic information and speaker identity. We note that the decoupling of emotional features from other speech information (such as speaker, content, etc.) is the key to achieving remarkable performance. Some recent attempts  about speech representation decoupling on the neutral speech can not work well on the emotional speech, due to the more complex acoustic properties involved in the latter.
To address this problem, here we propose a novel Source-Filter based Emotional VC model (SFEVC) to achieve proper filtering speaker-independent emotion features from both the timbre and pitch features.
Our SFEVC model consists of multi-channel encoders, emotion separate encoders and decoder. Note that all encoder modules adopt a designed information bottlenecks auto-encoder.
Additionally, to further improve the conversion quality for various emotions, a novel two-stage training strategy based on the 2D Valence-Arousal (VA) space was proposed. Experimental results show that the proposed SFEVC along with two-stage training strategy outperforms all baselines and achieves the state-of-the-art performance in speaker-independent emotional VC with nonparallel data.

\end{abstract}

\section{Introduction}
\IEEEPARstart{E}{motional} voice conversion (VC) is a useful speech processing technique for changing the emotional states of a speech utterance while retaining its linguistic information and speaker identity. It can be applied in various domains, such as virtual assistants, call centers, and audiobook narration \cite{aihara2014preliminary,mori2006emotional,krivokapic2013rhythm,raitio2015phase}, etc. 

\begin{figure}[htp]
	\centering
	\includegraphics[width=1\columnwidth]{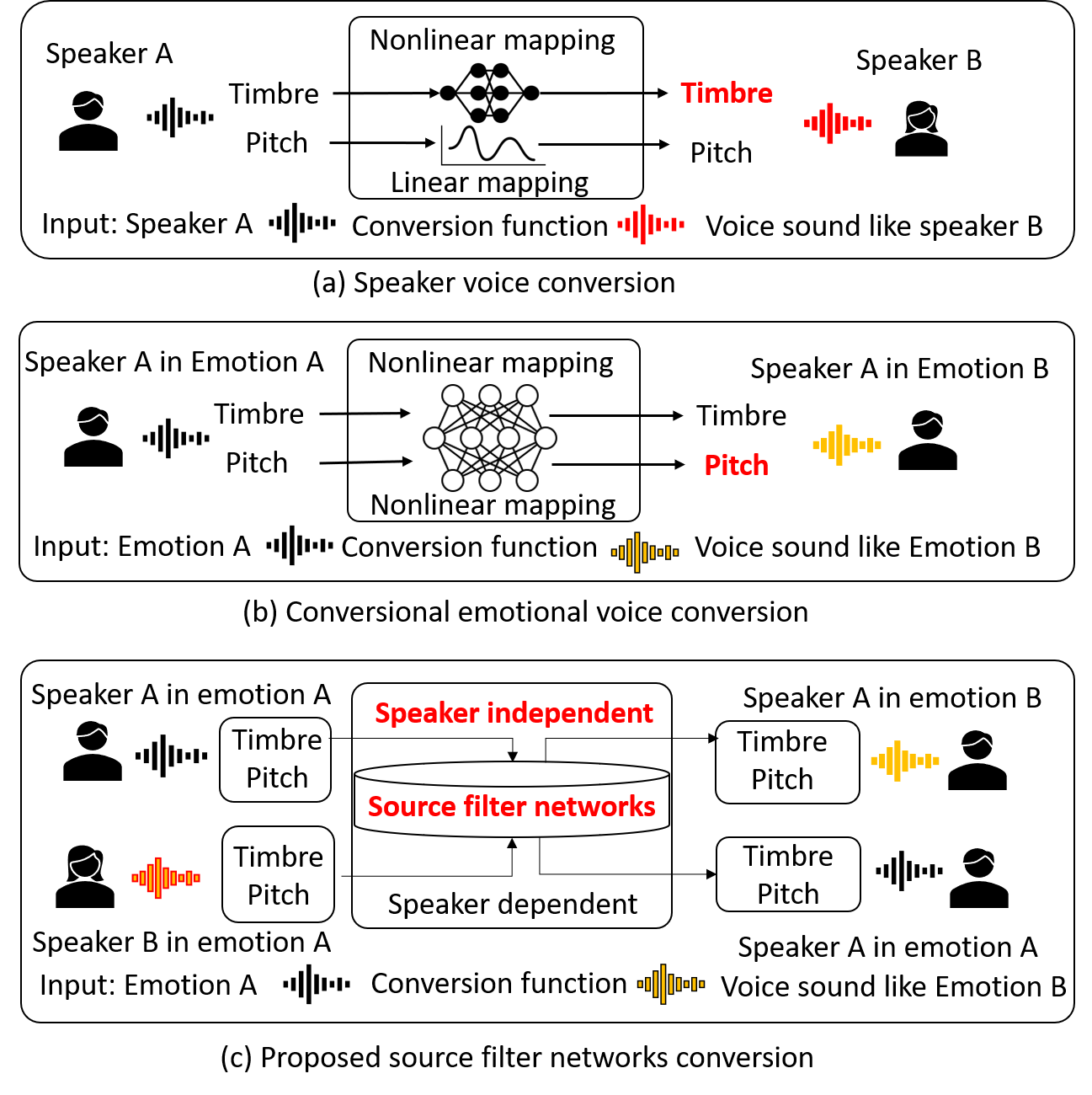}
	\centering
	\caption{Difference between (a) speaker voice conversion, (b) conventional emotional voice conversion, and (c) proposed source filter networks method.}	
	\label{fig:f1}
\end{figure}

Originally motivated by the traditional speaker VC, the study of emotional VC has recently attracted wide attention in the field of speech processing. However, emotional VC and traditional speaker VC differ in many ways. As shown in Fig. 1 (a), in the speaker VC tasks, e.g. the one-to-one conversion \cite{nakashika2013voice,miyoshi2017voice}, or the many to many conversion \cite{kameoka2018stargan,lee2020many}, the main purpose is to convert the source speaker's voice to sound like another speaker. In the emotional VC, we only convert the prosody of the voice to that of a target emotion such as anger or happiness, of the same person rather than another speaker. In a nutshell, for voice conversions, the traditional speaker VC aims to change the speaker's identity, whereas the emotional VC tries to convert the emotional states of the same speaker.

Traditional VC research includes modeling the timbre features mapping with statistical methods such as the Gaussian mixture model (GMM) ~\cite{toda2007voice}~\cite{helander2010voice} and non-negative matrix factorization (NMF)~\cite{takashima2012exemplar,fu2016joint}. Recent deep learning approaches such as deep neural network (DNN) \cite{lecun2015deep}, long short-term memory network (LSTM) \cite{sak2014long}, and generative adversarial networks (GANs) \cite{goodfellow2014generative} have achieved remarkable performance in the traditional VC \cite{xie2016kl,sun2015voice,kameoka2018stargan}. As a consequence, emotional VC has also developed in this direction. Early studies on emotional VC handle both timbre (spectrum) and pitch (F0) conversion with GMM \cite{aihara2012gmm, kawanami2003gmm,hsia2007conversion}. Some deep learning-based emotional VC models, such as DNN, RNN, and GANs have shown the effects on emotional VC. For example, Luo \textit{et al.} \cite{luo2016emotional} increased the dimension of the F0 features and applied the DNN model in the emotional VC. Moreover, for the data augmentation, they used the continuous wavelet transform (CWT) to analyze the F0 features \cite{luo2017emotional} and improve their work using the dual-supervised GANs models to do the training \cite{luo2019emotional}. Ming \textit{et al.} \cite{ming2016deep} applied the LSTM models in the emotional VC, and Kun \textit{et al.} \cite{zhou2020transforming} used the unsupervised cycleGAN model to do the nonparallel conversion in the emotional VC. These works have made a great contribution to the development of emotional VC.

However, as shown in Fig. 1 (b), conventional emotional VC methods just applied the nonlinear mapping model to the pitch (F0) conversion, which is similar to the normal speaker VC models used in the timbre features conversion as in Fig. 1 (a). In our primary experiments, we have observed that the speaker similarity will reduce the effectiveness of the emotional VC. We believe that this performance degradation is probably due to the ``noise'' in the emotional features. The common emotional features converted by traditional VC models include not only the emotion but also other information (e.g. target speaker identity) altogether, resulting in a converted emotional speech that does not retain the source speaker identity well. Nonetheless, as we hear from different speakers, even in different languages, we can easily recognize their emotions from the speech. Motivated by this, we find that there are speaker-independent emotion codes, which can be extracted from different speakers and languages. Therefore, in this work, we focus on disentangling the emotion feature from the other acoustic features and achieve emotional VC effectively. As shown in Fig. 1 (c), to address this problem, we propose a source-filter emotional VC networks (\textit{SFEVC}) to decouple the speaker-independent emotion code from other acoustic information of different speakers, but keep the speaker-dependent features of the source speaker unchanged.

The source-filter model \cite{taylor2010contribution} represents the speech production process by separating the excitation and the resonance phenomena in the vocal tract, where the source corresponds to the glottal excitation and the filter corresponds to the vocal tract. 
This model assumes that these two phenomena are completely decoupled. We note that the use of the source-filter model in emotional VC deserves further exploration, which will be to shown as follows in this paper.

In our SFEVC model, the source-filter network is based on the encoder-encoder-decoder architecture, which consists of multi-channel encoders, emotion-separate encoders, and the decoder. All encoders are applied with the designed information bottleneck auto-encoders, which can filter the specified features from the emotional speech. For the multi-channel encoders, they can disentangle the content from the acoustics features (timbre and prosody). However, these speaker-dependent acoustics features are still mixed with the emotion features. Therefore, the emotion-separate encoders are focusing on separating the speaker-independent emotion features from these speaker-dependent acoustic features. Finally, the decoder takes the speaker-independent emotion code as input to convert the speaker's emotion without distorting the speaker-dependent acoustic features.

Moreover, according to the emotion studies \cite{kensinger2004remembering, warriner2013norms, kuperman2014emotion}, psychological emotion labels can be typically divided into discrete emotion states (angry, happy, neutral, and so on) or dimensional continuous emotion space (valence-arousal (VA) space) \cite{adolphs2002recognizing}. As indicated in the emotion research, valence (how positive or negative is an emotion) and arousal (power of the activation of the emotion) constitute popular and effective representations for affecting the emotion. Using the VA space over the two emotion dimensions is considered to be more general than the use of discrete states in solving the speech problem. For example, the pitch feature of audio, one of the most reliable features, can be seen as the index of arousal \cite{johnstone2000vocal}, while it can be difficult to represent the valence features. Therefore, the happiness and sadness should be far away from each other in the VA space, while sadness and anger might be closer instead. To further improve the conversion performance in terms of emotion expressiveness, a novel two-stage training strategy, which leveraging the relationship between discrete emotion classes and VA space, was proposed to train our SFEVC model more effectively.

\begin{figure*}[htp]
  \centering
  \includegraphics[width=1\linewidth]{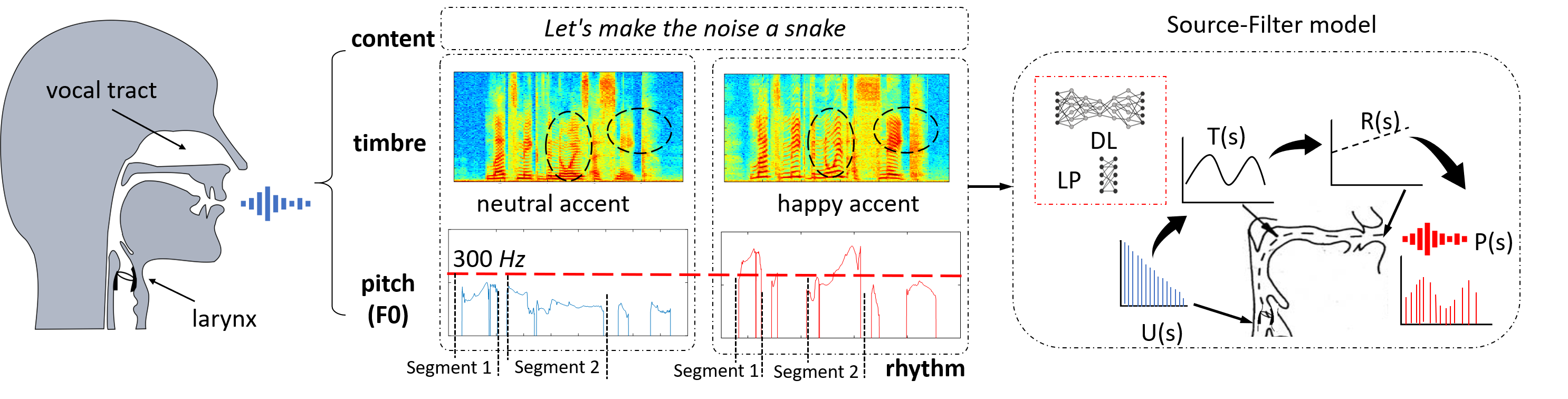}
  \caption{Emotional speech information and source-filter networks}
  \label{figure:speech information}
\end{figure*}

To validate our proposed model, we conduct the experiments on multiple emotional voice dataset. Note that we introduce a new sales conversation corpus with a new high tension emotion, denoted as ``\textbf{H}igh \textbf{T}ension \textbf{E}motion dataset (\textbf{HTE})'', that will contribute to enrich the development of VC research community.

In summary, the main contributions of this work are summarized as follows:
1) We point out that there are speaker-independent emotion codes, which can be extracted from different speakers and languages and help improve emotional VC performance.
2) We introduce a novel emotional VC paradigm based on the source-filter model to disentangle the emotion feature from other acoustic features. 
3) A novel two-stage training strategy based on the 2D valence-arousal space was proposed for an effective training procedure.
4) To validate the effectiveness of our proposed model, we include a new and challenging high tension emotion dataset, which has more complex emotions and can enrich the types of emotions for open-source emotional speech data.
To our best knowledge, this is the first study of applying the source-filter model in the emotional VC literature.

	
This paper is organized as follows. The background of emotional speech and source-filter theory are briefly reviewed in Section~\ref{Sec.related}. Emotion voice data analysis is provided in Section~\ref{Sec:data}. Section~\ref{Sec:proposed} presents our proposed SFEVC method. Section~\ref{Sec:exe} gives the details of experimental evaluations, and Section~\ref{Sec:con} concludes the paper.


\section{Background}
\label{Sec.related}
We provide here a brief primer on the emotional speech, the source-filter model and the emotional valence-arousal space theory.
\subsection{Information in Emotional Speech}
The left part of Fig. \ref{figure:speech information} shows the process of speech generation. The speech signal is first generated by an excitation signal in the larynx. Then the generated signal is modulated by resonance through the vocal tract (guttural, oral, and nasal cavities) which acts as a filter. The speech signal includes four main speech information components: language content, timbre, pitch, and rhythm. The emotional features are included in these components and can be represented in different ways. 

\textbf{Timbre} is reflected by the formant, which is the broad spectral maximum that results from an acoustic resonance of the human vocal tract. The timbre can represent the tone color or unique quality of a sound, which helps us instantly identify and classify sound sources, such as individual people or musical instruments. In the emotional voice, the high arousal voice, such as happy or angry voice, tends to sound sharper and brighter than the low arousal voice or neutral voice \cite{pittam1990long}. As shown in the middle of Fig. \ref{figure:speech information}, the timbre features of neutral voice and happy voice, we can see that the happy voice has a sharper spectrogram in some words, and a higher formant frequency range at the end utterance than the neutral speech, which indicates a brighter voice. Thus, it is also necessary to extract the emotional information from the timbre features.

\textbf{Pitch} is an important parameter in emotional speech processing systems. Modulated pitch is generated by the larynx and modulated primarily by fine changes in the tension of the vocal folds. The ability to voluntarily and flexibly control pitch patterns, in the context of vocal learning, is unique to humans among primates \cite{dichter2018control}, while it has been previously thought that this ability was due to anatomical differences in the larynx \cite{belyk2017origins}. As shown in pitch figures of Fig. \ref{figure:speech information}, the happy voice has a higher frequency than the neutral voice. It has been proved that the higher pitch, increased intensity, and faster rate were associated with more excited and positive emotions in speech \cite{ma2015human}. Thus, in this research, we will transform and control the pitch of voice by the designed emotion auto-encoder, for flexible emotion conversion. However, the pitch contour also entangles the rhythm information and speaker identity. For example, female speakers tend to have a higher pitch range than male speakers, which indicates certain speaker identity information. Moreover, since each nonzero segment of the pitch contour represents a voiced segment, the length of each voiced segment indicates how fast the speaker speaks, which can be represented as the rhythm information.

\textbf{Content} belongs to the language model in speech research. In the content information, the phoneme is the basic unit in most languages. Each phoneme has a particular formant pattern. Thus different phonemes will be represented in different shapes in the spectrogram. As shown in the spectrogram in Fig. \ref{figure:speech information}, the spectrograms of the same content spoken in different emotions have a similar shape. In the emotional VC tasks, the linguistic information needs to be kept unchanged. Thus, the source-filter model needs to retain the content information while converting the emotion features. 

\textbf{Rhythm} is a recurring movement of sound or speech. It represents how fast the speaker utters each syllable or word. Regular recurrence of grouped stressed and unstressed, long and short, or high-pitched and low-pitched syllables goes in alternation. As shown in Fig. \ref{figure:speech information}, each pitch contour is divided into segments, which correspond to vowels of words, and the lengths of these segments reflect the rhythmic information. In the speech, different emotional information can be expressed at different speaking speeds, for instance, the happy and angry voice is usually faster than the sad and neutral voice.

\subsection{The Source-Filter Model}
The source-filter model simulates the speech production via two distinctive parts, i.e. the excitation in the larynx and the resonance in the vocal tract. This principle is illustrated in the right side of Fig. \ref{figure:speech information}. The vowel spectrum $P(s)$ is the product of the spectrum of the glottal source $U(s)$, the transfer function of the vocal tract $T(s)$ (filter), and the radiation characteristics $R(s)$. This model is also known as the “source-filter theory” of vowel production \cite{stevens1998acoustic}. An example of source-filter modeling is the Linear Prediction (LP) model \cite{hall1983time}, which uses the source-filter theory assuming that the speech is the output signal of a recursive digital filter when excitation is received at the input. In reality, the mechanism of the vocal fold is more complex, making this assumption over-simplistic. There exist other source-filter models that use multi-layer deep learning (DL) models to deal with the complex excitation signals composed of deterministic and stochastic components \cite{wang2019neural}. Some attempts in traditional speaker VC also applied source-filter theory on feature decoupling, but are mostly focused on simulating the vocal tract. Hence they only convert the timbre feature of the voice \cite{nakashika2014high,qian2019autovc} or separate speaker-dependent timbre and prosody from the content \cite{qian2020unsupervised}, while the speaker-independent emotional information is still mixed in the prosody and timbre features. 
In the emotional VC, we propose a source-filter model that can decouple the speaker independent emotion information from both the timbre and pitch features.

\subsection{Valence-Arousal}
Representing human emotions has been a basic research topic in psychology. The most frequently used emotion representation is the categorical one, comprising several basic categories such as anger, disgust, fear, happiness, sadness, surprise and neutral, etc. It is, however, the dimensional emotion representation \cite{whissell1989dictionary} that is more appropriate to depict subtleties. Most of the dimensional models classify affective states in two dimensions, i.e. `Valence' and `Arousal'. As indicated in the emotion research, the 2D VA space provides a popular and effective representation for affecting emotions. For example, the pitch feature of audio, one of the most reliable features, can be seen as the index of arousal \cite{johnstone2000vocal}, while the facial expression can be used as the valence. Thus, we can improve the training effectiveness for the emotional VC via the two-stage learning pipeline, based on the relationships between emotions' valence-arousal (VA) spaces.

\section{Data Analysis}
\label{Sec:data}
\begin{figure}[tp]
\centering
   \begin{subfigure}[b]{1.0\columnwidth}
   \includegraphics[width=1.0\textwidth]{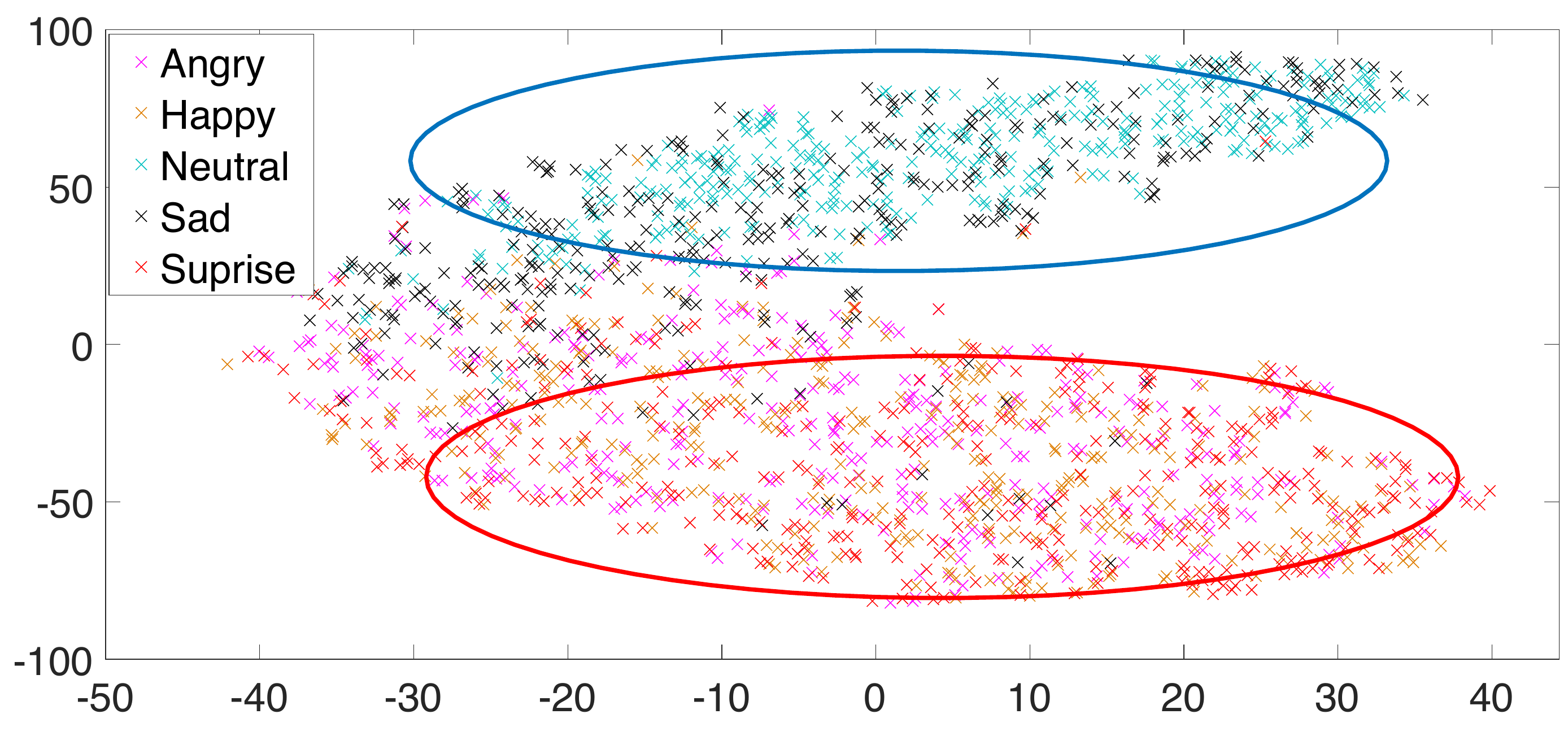}
   \caption{}
   \label{fig:Ng1} 
\end{subfigure}
\begin{subfigure}[b]{1.0\columnwidth}
   \includegraphics[width=1\columnwidth]{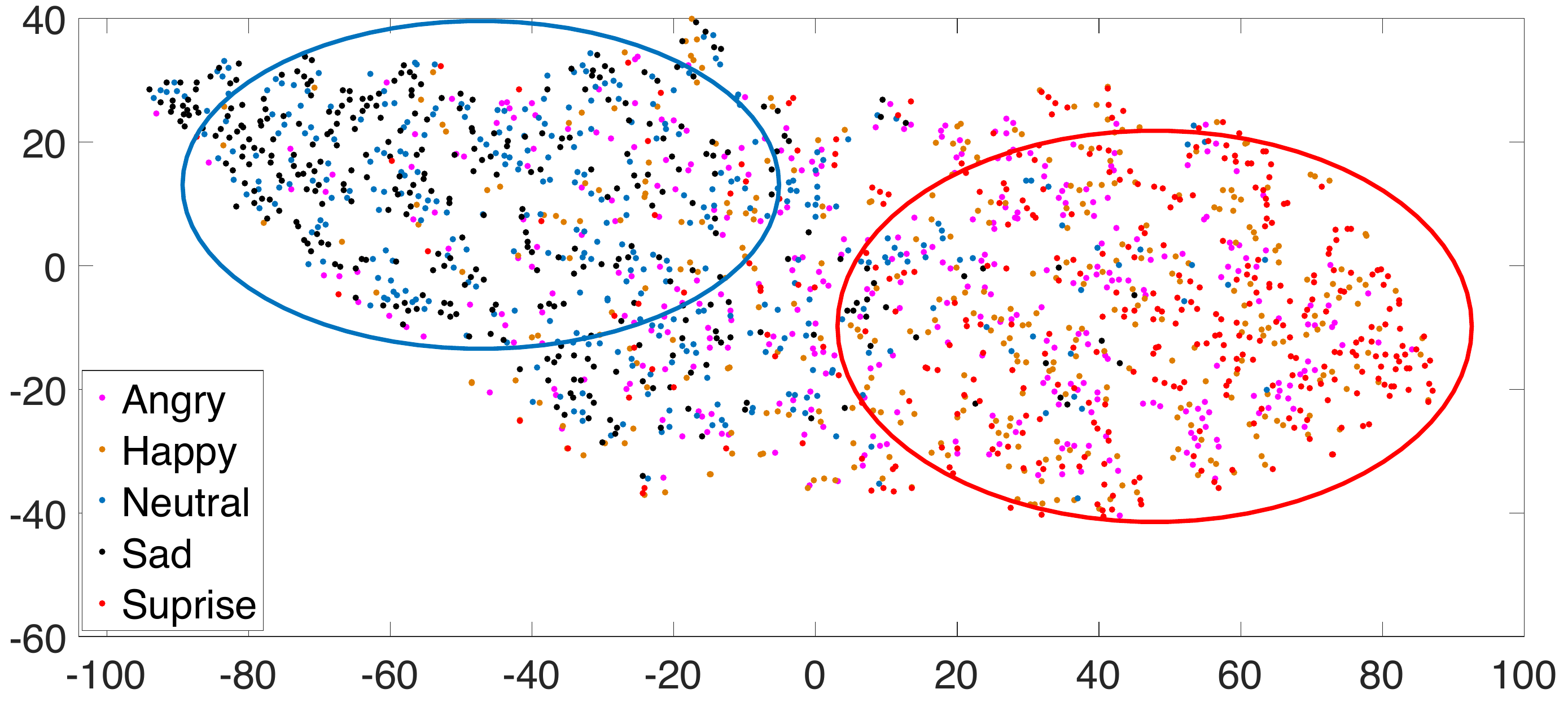}
   \caption{}
   \label{fig:Ng2}
\end{subfigure}
\caption{(a) t-SNE visualization of F0 features from 350 sentences with the same content spoken in Chinese. (b) t-SNE visualization of F0 features from 350 sentences with the same language content spoken in English.}
\label{fig:f3}
\end{figure}

In this chapter, we first provide the speech data analysis for the most important pitch features F0 in the emotional VC. We apply the t-distributed Stochastic Neighbor Embedding (t-SNE) \cite{van2008visualizing} to reduce the dimensionality of the emotional features of the different emotional speech and to plot them in a two-dimensional space. Fig. \ref{fig:f3} (a) and Fig. \ref{fig:f3} (b) represent the t-SNE separation by the F0 features, which are the main emotion representation in speech. The instances are marked per emotions in two different languages. As shown in the figures, the neutral is mostly mixed with the sad emotion, while the happy, surprised and angry mixed together. The low arousal emotions (sad and neutral) can be easily separated from high arousal emotions (happy, angry and surprised), while it is difficult to separate emotions in the same arousal group, no matter how different their valence it is. Thus, it is easier to represent the emotion's arousal features  with the prosody feature of speech, rather than the valence features. Therefore, when converting the emotions, we need to train the emotional source-filter network with different stages based on the 2D VA space.

\begin{figure}[tp]
	\centering
	\vspace{-5mm}
	\includegraphics[width=1\columnwidth]{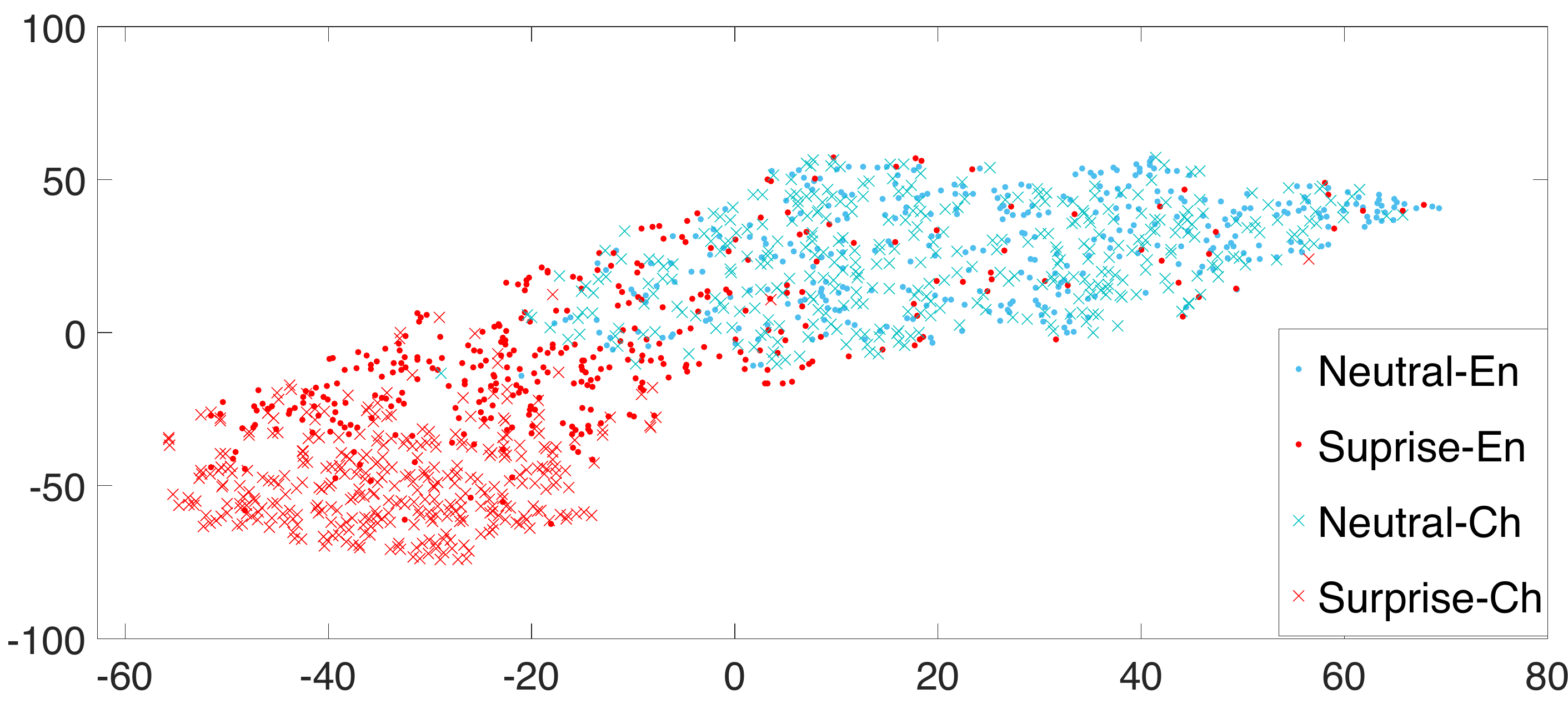}
	\centering
	\caption{t-SNE visualization of F0 features of two emotions spoken in different languages.}
	\label{fig:f4}
\end{figure}

As shown in Fig. \ref{fig:f4}, we also apply the t-SNE to plot the emotion features extracted from different speakers in different languages. We color-coded the instances according to emotions and the different shapes of the data points show the different speakers in different languages. We can expect to see clusters forming based on emotions instead of languages, which indicates that the emotion is a speaker-independent feature, even in the different languages. Motivated by this, we can use different speakers' emotional speech to extract the speaker-independent emotion code as shown in Fig 1 (c). Then, we use the different emotion speech by the same speaker to train the source-filter networks to learn how to filter out the emotion feature but keep the speaker-dependent pitch and timbre features unchanged.

\begin{figure*}[htp]
  \centering
  \includegraphics[width=1\linewidth]{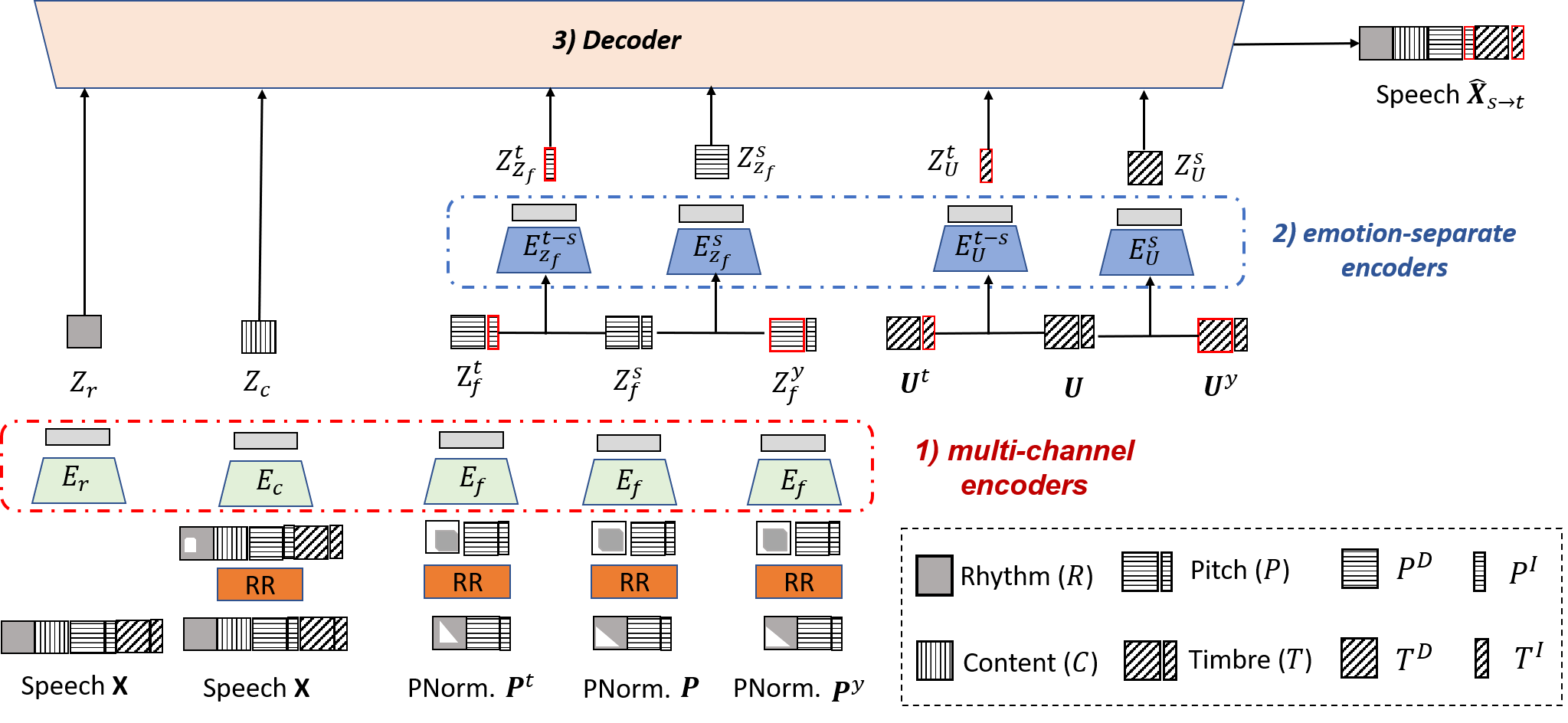}
  \caption{The framework of proposed SFEVC. ‘RR’ denotes random resampling. ‘PNorm.’ is short for the normalized pitch contour. $\boldsymbol{P}$, $\boldsymbol{P^{t}}$, and $\boldsymbol{P^{y}}$ represents the normalized pitch contour of source emotion, target emotion and source emotion of another speaker ($\boldsymbol{Y}$), respectively. $\boldsymbol{U}$, $\boldsymbol{U}^{t}$, and $\boldsymbol{U}^{y}$ represents the input timbre features of source emotion speech, target emotion speech and source emotion speech of another speaker, respectively. $(E)$s and $(Z)$s represent the encoders and their codes, respectively. Pitch feature ($P$) consists of the speaker-dependent pitch feature ($P^D$) and the speaker-independent pitch feature ($P^I$). Timbre feature ($T$) consists of speaker-dependent timbre feature ($T^D$) and speaker-independent timbre feature ($T^I$). Some rhythm blocks have some holes in them, which represents that a portion of the rhythm information is lost. The grey block at the tip of the encoders denotes the information bottleneck.}
  \label{figure:network_structure}
\end{figure*}

\section{Source-Filter Emotional Voice Conversion}
\label{Sec:proposed}
\subsection{The SFEVC Framework}
As shown in Fig. 5, our proposed SFEVC model consists of 1) multi-channel encoders, 2) emotion-separate encoders and 3) decoder. In the multi-channel encoders, each channel has a different and carefully crafted information bottleneck design to decompose speech into content, prosody, and rhythm, separately. 
In the emotion-separate encoders, the speaker-independent emotion information can be separated from the source and target pitch codes decomposed from the multi-channel encoders and the timbre features from the source and target speech. The decoder aims to take the decoupled features as input to generate the target speech spectrogram features. We will introduce all modules in order.

\subsubsection{Multi-Channel-Encoders}

We apply the multi-channel-encoder with three encoder channels ${E_{r}, E_{c},E_{f}}$. Here $E_{r}$ denotes the rhythm encoder, $E_{c}$ denotes the content encoder, and $E_{f}$ denotes the pitch encoder. Each channel has a different, carefully crafted information bottleneck design, which is similar to AutoVC \cite{qian2019autovc}.

\subsection{Training procedure based on valence-arousal features}
\begin{figure}[t]
  \centering
  \includegraphics[width=1.0\linewidth]{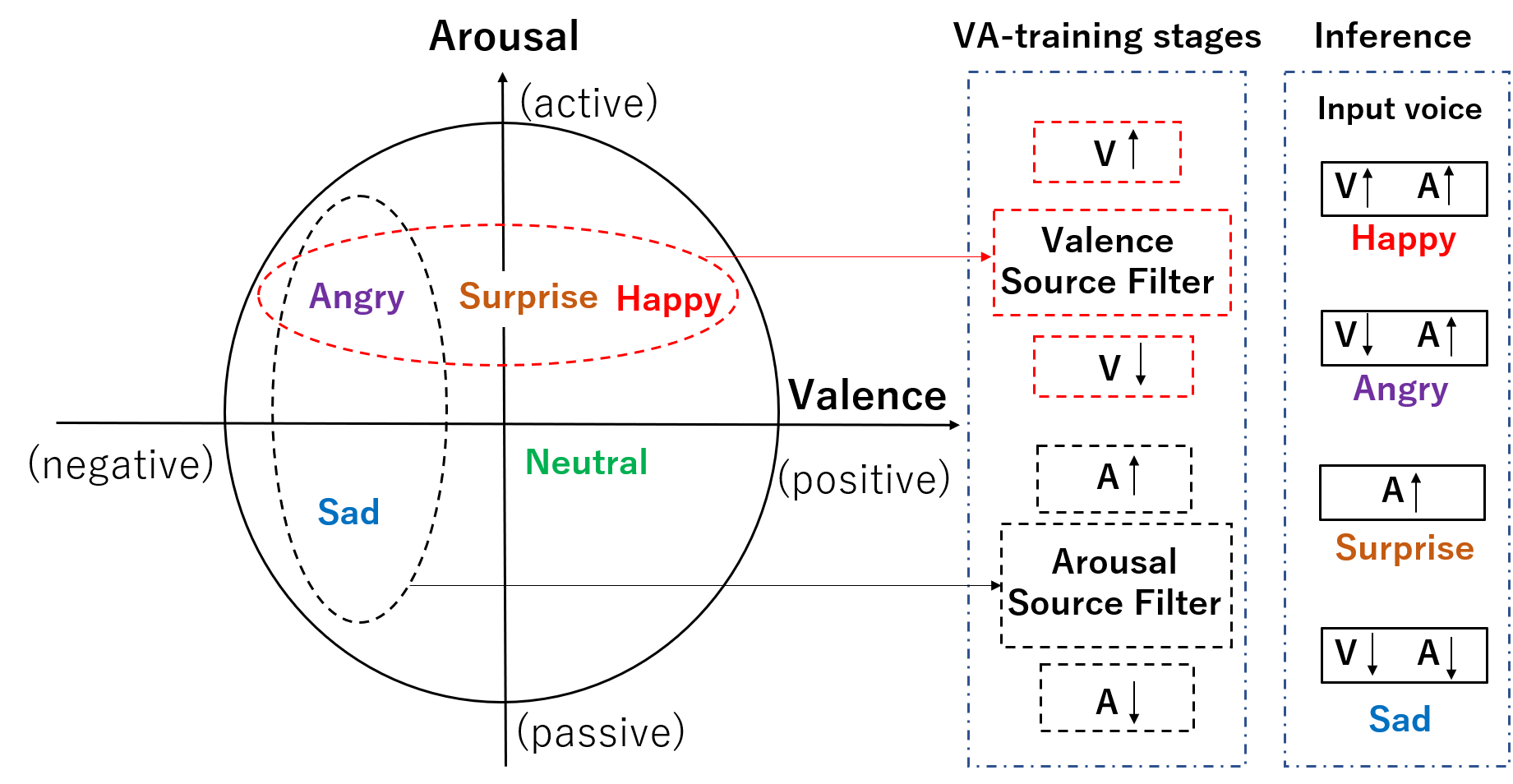}
  \caption{Training staged based on arousal valence features.}
  \label{fig:AV}
\end{figure}

The inputs of the multi-channel encoders are speech $\boldsymbol{X}$ and normalized pitch contours (PNorm.$\boldsymbol{P}$, PNorm.$\boldsymbol{P^{t}}$, and PNorm.$\boldsymbol{P^{y}}$). As the normalized pitch contours are normalized to have the same mean and variance across the same speakers, so the normalized pitch contours only contain the pitch information and rhythm information, but no content information and timbre information.

The output of the encoders called codes $\boldsymbol{Z} =\left \{ {\boldsymbol{Z}_{r},\boldsymbol{Z}_{c},\boldsymbol{Z}_{f}^{s},\boldsymbol{Z}_{f}^{t},\boldsymbol{Z}_{f}^{y}} \right \}$, where $\boldsymbol{Z}_{r}$, $\boldsymbol{Z}_{c}$, $\boldsymbol{Z}_{f}^{s}$, $\boldsymbol{Z}_{f}^{t}$, and $\boldsymbol{Z}_{f}^{y}$ denotes rhythm code, content code,pitch code of source emotion, pitch code of target emotion, pitch code of another speaker, respectively. The codes can be expressed as follows:
\begin{equation}
\begin{split}
    \label{equation:1}
    \boldsymbol{Z}_{r}=E_{r}(\boldsymbol{X}),
    \boldsymbol{Z}_{c}=E_{c}(A(\boldsymbol{X})),\\  \boldsymbol{Z}_{f}^{s}=E_{f}(A(\boldsymbol{P})),
    \boldsymbol{Z}_{f}^{t}=E_{f}(A(\boldsymbol{P^{t}})),
    \boldsymbol{Z}_{f}^{y}=E_{f}(A(\boldsymbol{P^{y}})),
\end{split}
\end{equation} 
where $A(\cdot)$ denotes the random resampling (RR) operation. As shown in Fig. 5, we applied RR operation for content encoder and pitch encoder, but not rhythm encoder. Because the RR operation divides the input into segments of random lengths and randomly stretch or squeeze each segment along the time dimension. Therefore, it can be used as an information bottleneck to filter out rhythm information.

\subsubsection{Emotion-Separate Encoders}
As described above, in the emotional VC tasks, we regard the emotion information as speaker-independent features and they are mixed in the timbre and prosody features. Thus, in the emotional VC, we need to decouple the speaker-independent emotion feature from the timbre features $(\boldsymbol{U},\boldsymbol{U}^{t})$ and pitch codes $(\boldsymbol{Z}_{f}^{s},\boldsymbol{Z}_{f}^{t})$ extracted from the multi-channel encodes, while keeping the speaker dependent features unchanged. In the emotion-separate encoders, $E_{\boldsymbol{U}}^{s}$ and $E_{\boldsymbol{Z}_{f}}^{s}$ are the speaker dependent features encoders for timbre and pitch features, respectively. $E_{\boldsymbol{U}}^{t-s}$ and $E_{\boldsymbol{Z}_{f}}^{t-s}$ represent the speaker-independent emotion encoders for timbre and pitch features, respectively. The emotion-separate encoders can be expressed as follows:
\begin{equation}
\begin{split}
    \label{equation:3}
     \boldsymbol{Z}_{U}^{s}= E_{\boldsymbol{U}}^{s}(\boldsymbol{U},\boldsymbol{U^y}),\\ \boldsymbol{Z}_{U}^{t}= E_{\boldsymbol{U}}^{t-s}(\boldsymbol{U^{t},U}),\\
    \boldsymbol{Z}_{\boldsymbol{Z}_{f}}^{s}=    E_{\boldsymbol{Z}_{f}}^{s}(\boldsymbol{Z}_{f}^{s},\boldsymbol{Z}_{f}^{y}),\\ \boldsymbol{Z}_{\boldsymbol{Z}_{f}}^{t}=    E_{\boldsymbol{Z}_{f}}^{t-s}   (\boldsymbol{Z}_{f}^{t},\boldsymbol{Z}_{f}^{s}).
\end{split}
\end{equation} 

 To let the $E_{\boldsymbol{U}}^{t-s}$ extract speaker-independent emotion features from timbre features, the inputs are timbre features $(\boldsymbol{U^{t},U})$ of source and target emotion speech ($s$ and $t$) from the same speaker ($\boldsymbol{X}$). To let $E_{\boldsymbol{Z}_{f}}^{t-s}$ extract speaker-independent emotion features from pitch features, the inputs are $(\boldsymbol{Z}_{f}^{s},\boldsymbol{Z}_{f}^{t})$ from the multi-channel encoders by the same speaker. To extract the speaker-dependent features of the timbre and pitch features by encoders $E_{\boldsymbol{U}}^{s}$ and $E_{\boldsymbol{Z}_{f}}^{s}$, the timbre features $\boldsymbol{U},\boldsymbol{U^y}$ and pitch codes $\boldsymbol{Z}_{f}^{s},\boldsymbol{Z}_{f}^{y}$ of the source emotion speech by different speakers ($\boldsymbol{X}$ and $\boldsymbol{Y}$) are as the inputs for training encoders. 

The speaker-dependent emotion codes extracted from timbre and prosody features can be represented as $\boldsymbol{Z}_{U}^{s}$ and $\boldsymbol{Z}_{\boldsymbol{Z}_{f}}^{s}$, while the speaker-independent emotion codes extract from the timbre and prosody can be represented as $\boldsymbol{Z}_{U}^{t}$ and $\boldsymbol{Z}_{\boldsymbol{Z}_{f}}^{t}$.

\subsubsection{Decoder}
The decoder takes $\boldsymbol{Z}$ as its inputs and produces a speech spectrogram $ \hat{\boldsymbol{X}}_{s\rightarrow t}$ as output. Now we want to convert the voice's speaker independent emotion from the source emotion $\boldsymbol{X}$ to the target emotion $\boldsymbol{X}^t$ but keep the speaker identity unchanged. The converter should have the following desirable property:
\begin{equation}
\begin{split}
    \label{equation:2}
    \hat{\boldsymbol{X}}_{s\rightarrow t}= D(\boldsymbol{Z}_{r},\boldsymbol{Z}_{c},\boldsymbol{Z}_{Z_{f}}^{t},\boldsymbol{Z}_{Z_{f}}^{s},\boldsymbol{Z}_{U}^{t},\boldsymbol{Z}_{U}^{s}).
\end{split}
\end{equation} 
where, $\boldsymbol{Z}_{r}$ and $\boldsymbol{Z}_{c}$ are the source speech rhythm and content, $\boldsymbol{Z}_{Z_{f}}^{s}$ and  $\boldsymbol{Z}_{U}^{s}$ are the speaker dependent pitch and timbre from source speech. $\boldsymbol{Z}_{Z_{f}}^{t}$ and  $\boldsymbol{Z}_{U}^{t}$ are the speaker independent pitch and timbre from target emotion speech.

During training, the output of the decoder tries to reconstruct the input spectrogram:
\begin{equation}
    \label{equation:4}
    \underset{\theta}{\mathrm{min}}\ \mathbb{E} \left [\left \| \hat{\boldsymbol{X}}_{s\rightarrow t}-\boldsymbol{X} \right \|  \right] _{2}^{2} 
\end{equation} 
where $\theta$ denotes all the trainable parameters. It has been proved that if all the information bottlenecks are appropriately set and the network representation power is sufficient, a minimizer of equation \ref{equation:4} will satisfy the multi-channel encoders and emotion-separate encoders, separately. More details can be found in appendix.

\subsection{Inference Method}
In this section, we explain why SFEVC can achieve speech decomposition by multi-channel encoders and filter the speaker-independent emotion features from the timbre and pitch features by emotion-separate encoders. 
The theory of how multi-channel encoders achieve speech decomposition is similar to the other source filter networks applied in the speaker VC tasks \cite{qian2019autovc,qian2020unsupervised}. When passing through the random resampling (RR) operation, a random portion of the rhythm block is wiped, but the other blocks remain intact. Thus, when the speech and pitch features pass through the random resampling operation, the rhythm blocks of them are missing information. Rhythm encoder $E_{r}\left (\cdot  \right)$ is the only encoder that has access to the complete rhythm information, So if $E_{r}\left (\cdot  \right)$ is forced to lose some information by its information bottleneck, it will prioritize removing the content, pitch, and timbre. For the same reason, the $E_{r}\left (\cdot  \right)$ only encodes rhythm, then the content encoder $E_{c}\left (\cdot  \right)$ becomes the only encoder that can encode all the content information, thus, it will keep the content and removing the other features by the designed bottlenecks layers. Finally, with $E_{r}\left (\cdot  \right)$ encoding only rhythm and $E_{c}\left (\cdot  \right)$ encoding only content, the pitch encoder $E_{f}\left (\cdot  \right)$ must encode the pitch information.

Different from the conventional source-filter networks for normal speaker voice conversion, where the speaker's identity is directly fed to the decoder, in the emotional VC, we need to convert the emotion of voice but keep the speaker identity unchanged. Thus, we propose the emotion-separate encoders to encode the timbre features $(\boldsymbol{U},\boldsymbol{U}^{t})$ and the pitch codes $(\boldsymbol{Z}_{f}^{t},\boldsymbol{Z}_{f}^{s})$. As shown in the emotion-separate encoders, the speaker-independent encoders ($E_{\boldsymbol{U}}^{t-s}$ and $E_{\boldsymbol{Z}_{f}}^{t-s}$) encode the paired source and target emotion by the same speaker $\boldsymbol{X}$. Thus, these encoders are only embedded with the emotion codes without the speaker's identity. For the speaker-dependent encoders ($E_{\boldsymbol{U}}^{s}$ and $E_{\boldsymbol{Z}_{f}}^{s}$), the inputs are the same emotional voices spoken by different speakers, which can be embedded with the speaker's identity in the pitch and timbre features. In general, through these emotion-separate encoders, speaker-dependent and speaker-independent features can be decoupled from the timbre and pitch features.

As described in the data analyses in Section III, sad and neutral are mixed, while happy, angry, and surprise emotions are mixed when using the t-SNE for clustering with pitch features. As shown in Fig. \ref{fig:AV}, in the emotion wheel, neutral and sadness belong to middle or low arousal emotion features, while happy, surprise, and angry are the high arousal features. It indicates that it is efficient to convert the emotion in different arousal, for example, from (neutral and sad) to high arousal (happy, angry, and surprise) using the pitch features. However, it is difficult to convert the valence features using pitch features. Therefore, in our training pipeline, we train the emotion source-filter using the separated stages based on the valence-arousal features.

For training the arousal source filter, we used the emotion dataset in a similar valence area, but the arousal is different. For instance, we train the conversion function from angry to happy to get the arousal conversion code, which is focused on converting the low arousal to high arousal ($A\uparrow$), and their inverse conversion (happy to angry) can convert the high arousal features to low arousal, which can be represented as $A\downarrow$. Similar processing to the valence convert training function, we can get the valence source filters as ($V\uparrow$) which can filter the low valence features, and ($V\downarrow$), which can filter the high valence features. 

As our goal is to convert the neutral voice to the common emotion voice, therefore, in the inference stage, our input is the neutral voice. As shown in the inference part in Fig. \ref{fig:AV}, when we want to convert neutral voice to happy voice, we need to filter the neutral voice by the ($A\uparrow$) and ($V\uparrow$), which increases its arousal and valence, respectively. To convert to the angry voice, it needs to decrease the valence and increase the arousal using the ($A\downarrow$) and ($V\uparrow$). For the sad voice, ($A\downarrow$) and ($V\downarrow$) can be used.


\section{Experiments}
\label{Sec:exe}
\subsection{Dataset and Experimental Settings}

All experiments are conducted on four datasets including ATR~\cite{kawanami2003gmm}, JUST corpus~\cite{sonobe2017jsut}, ESD~\cite{zhou2021seen} and our new HTE datset.

\begin{itemize}
	\item\textbf{ATR~\cite{kawanami2003gmm}:} In the database, 50 sentences from the ATR Japanese phonetically balanced text set were used in the experiments. These 50 sentences are designed to include a minimum phoneme set of Japanese. All the texts were read by two female professional narrators with neutral, angry, happy, and sad voices.
	\item\textbf{JUST corpus~\cite{sonobe2017jsut}:} In the JUST corpus, 100 sentences are recorded by three female professional narrators with neutral, angry, happy, and sad voices.
    \item\textbf{ESD~\cite{zhou2021seen}:} The dataset consists of 350 parallel utterances with an average duration of 2.9 seconds spoken by 10 native English and 10 native Mandarin speakers. For each language, the dataset consists of 5 male and 5 female speakers in five emotions (happy, sad, neutral, angry, and surprise). 
    \item \textbf{HTE:} We introduced the HTE \footnote{\textbf{High tension emotion dataset (HTE):} anonymous} dataset for real world emotional VC testing evaluation. The dataset consists of 100 parallel utterances including 50 sales conversation utterances and 50 phonetically balanced sentences with an average duration of 5 seconds spoken by 6 native Japanese voice actors (3 males and 3 females). Each sentence is spoken with two scenarios: 1) acting as a salesperson who speaks in a high tension emotional voice, and 2) acting like a normal person who speaks in a neutral voice. Speech data are sampled at 16 kHz rate with 16 bits resolution. 
\end{itemize}
All the speech data are sampled at 16 kHz rate with 16 bits resolution. We set these four datasets into the following: neutral to angry voice (N2An), neutral to sad voice (N2Sa), neutral to happy voice (N2Ha), neutral to surprise voice (N2Su), and neutral to high tension voice (N2Hi). We conducted evaluations with a five-fold cross-validation scheme and the performance is measured using average conversion evaluation.
Please refer to our online demo \footnote{\textbf{Demo page:} anonymous} to enjoy the speech samples.

\subsection{Experimental Settings}
 As described in Section~\ref{Sec.related}, the networks consist of the multi-channel encoders, the emotion-separate encoders, and the output decoder. All the encoders share a similar architecture, which consists of convolutional layers followed by group normalization \cite{wu2018group}. The gray blocks at the tip of the encoders shown in Fig. \ref{fig:f1} are the designed information bottlenecks which are a stack of BLSTM layers. They are applied after the output of the convolutional layers to reduce the feature dimension. By using the designed bottlenecks, the information of each channel can be passed through a downsampling operation to reduce the temporal dimension, producing hidden representations. Table \ref{table:tb1} shows the hyperparameter settings of multi-channel encoders, $E_{r}$, $E_{c}$ and $E_{f}$. Table \ref{table:tb2} shows the hyperparameter settings of second-level emotion-separate encoders ($E_{\boldsymbol{U}}^{s}$, $E_{\boldsymbol{U}}^{t-s}$, $E_{\boldsymbol{Z}_{f}}^{s}$ and $E_{\boldsymbol{Z}_{f}}^{t-s}$). The decoder first upsamples the hidden representation to restore the original sampling rate. 
Then all the representations are concatenated along the channel dimension and fed to a stack of three BLSTM layers \cite{sun2015voice} with an output linear layer to produce the final output. The spectrogrum and F0 features are converted back to the speech waveform using the same wavenet-vocoder as in \cite{shen2018natural}  on
the VCTK corpus.

\begin{table} [t]
    \caption{Hyperparameter settings of multi-channel encoders}
    \vspace{-5pt}
    \begin{tabular}{c c c c} 
       \toprule
            & $E_{r}$& $E_{c}$& $E_{f_{e}}$ \\ 
       \midrule
           Conv Layers & 1 & 3 & 1   \\ 
           Conv Dim & 128 & 512 & 128  \\ 
           Norm Groups & 8 & 32 & 8  \\ 
           BLSTM Layers & 1 & 2 & 1  \\ 
           BLSTM Dim & 1 & 8 & 16  \\ 
           Downsample Factor & 8 & 8 & 8 \\ 
       \bottomrule
    \end{tabular}
    \centering
    \label{table:tb1}
\end{table}

\begin{table} [t]
    \caption{\it Hyperparameter settings of emotion-separate encoders}
    \vspace{-5pt}
    \begin{tabular}{c c c c c} 
       \toprule
            & $E_{\boldsymbol{U}}^{s}$&  $E_{\boldsymbol{U}}^{t-s}$& $E_{\boldsymbol{Z}_{f}}^{s}$& $E_{\boldsymbol{Z}_{f}}^{t-s}$  \\ 
       \midrule
           Conv Layers & 3  & 3 & 1  & 3 \\ 
           Conv Dim & 512  & 256 & 128 &256 \\ 
           Norm Groups &32  & 8 & 8  & 16\\ 
           BLSTM Layers &2  & 1 & 1  & 1\\ 
           BLSTM Dim & 8 & 16 & 4  & 8 \\ 
           Downsample Factor & 8 & 8 & 8 &8 \\ 
       \bottomrule
    \end{tabular}
    \centering
    \label{table:tb2}
\end{table}

\begin{table}[tp]
	\caption{MCD results for the conversion of neutral voice to emotional voice.}
	\centering
	\setlength\tabcolsep{4.5pt}
	\begin{tabular}{c|c c c c c}
		\toprule
		\multicolumn{1}{c|}{} & \multicolumn{1}{c}{N2An} & N2Sa & N2Ha & N2Su & N2Hi \\
		\midrule
		Source &3.45 & 3.03 & 3.18 & 3.30 & 3.52  \\
		\hline
		DNNs+NNs & 3.12 &  2.61 &  2.75 & 3.11 & 3.23 \\
		Dual-SANs & 2.82 &  2.31 &  2.55 & \textbf{2.62} & 3.14 \\
		CycleGAN & 3.02 &  2.35 &  2.65 & 2.94 & 3.21\\
		SPEECHFLOW & 2.85 &  2.33 &  2.45 & 2.79 & 3.11 \\
		\hline
		SFEVC  & 2.83 &  2.30 &  2.48 & 2.72 & 2.95 \\
		SFEVC+VA & \textbf{2.78} &  \textbf{2.22} &  \textbf{2.36} & 2.65 & \textbf{2.91} \\
		\bottomrule
	\end{tabular}
	\label{tb:result2}
\end{table}

\begin{table}[tp]
	\caption{F0-RMSE results for the conversion of neutral voice to emotional voice.}
	\centering
	\setlength\tabcolsep{4.5pt}
	\begin{tabular}{c|c c c c c}
		\toprule
		\multicolumn{1}{c|}{} & \multicolumn{1}{c}{N2An} & N2Sa & N2Ha & N2Su & N2Hi \\
		\midrule
		Source & 76.3 & 55.1 & 83.3 & 68.5 & 79.9  \\
		\hline
		DNNs+NNs & 38.2 & 35.4 & 42.9 & 58.2 & 68.3   \\
		Dual-SANs & 21.3 & 21.8 & 21.8 & 34.5 & 49.9   \\
		CycleGAN & 28.1 & 23.1 & 23.3 & 38.5 & 44.3 \\
		SPEECHFLOW & 26.3 & 22.3 & 21.3 & 28.5 & 41.9 \\ \hline
		SFEVC  & 24.8 & 19.9 & 20.3 & 29.1 & 29.9 \\
		SFEVC+VA & \textbf{16.3} & \textbf{19.1} & \textbf{19.2} & \textbf{28.8} & \textbf{28.1}\\
		\bottomrule
	\end{tabular}
	\label{tb:result3}
\end{table}

\renewcommand{\thesubtable}{\alph{subtable}}
\begin{table}[!th]
	\centering
	\caption{ Results of classification for recorded voices[\%] and converted voices [\%].} \label{tb:result5}

	\begin{subtable}[t]{\columnwidth}
	\centering
	\setlength\tabcolsep{4pt}
	\caption{Recorded voice} \label{table:1a}
	\vspace{-3pt}
	\begin{tabular}{c|c c c c c c}
		\toprule
		\multicolumn{1}{c|}{Tar./Percept} & \multicolumn{1}{c}{Sad} & Angry & Surprise & High Tension & Happy &  Neutral \\
		\midrule
		Sad & \textbf{99} & 0 & 0 & 0  & 0 & 1 \\
		Angry & 0 & \textbf{96} & 0 & 2  & 0 & 2\\
		Surprise & 0 & 0 & \textbf{94} & 2  & 2 & 2 \\
		High tension & 0 & 2 & 2 & \textbf{90}  & 6 & 0 \\
		Happy & 0 & 0 & 2 & 6  & \textbf{92} & 0 \\
		Neutral & 0 & 1 & 0 & 2  & 2 & \textbf{95} \\
		\bottomrule
	\end{tabular}
	\end{subtable}

	\begin{subtable}[t]{\columnwidth}
	\centering
	\setlength\tabcolsep{4pt}
	\vspace{12pt}
	\caption{DBNs+NNs converted voice} \label{table:1a1}
	\vspace{-3pt}
		\begin{tabular}{c|c c c c c c}
		\toprule
		\multicolumn{1}{c|}{Tar./Percept} & \multicolumn{1}{c}{Sad} & Angry & Surprise & High Tension & Happy &  Neutral \\
		\midrule
		Sad & \textbf{50} & 2 & 3 & 0  & 0 & 45 \\
		Angry & 5 & \textbf{48} & 2 & 10  & 0 & 35\\
		Surprise & 8 & 12 & \textbf{44} & 15  & 1 & 20 \\
		High tension & 1 & 20 & 10 & \textbf{28}  & 30 & 11 \\
		Happy & 0 & 5 & 18 & 30  & \textbf{33} & 14 \\
		\bottomrule
	\end{tabular}
	\end{subtable}

	\begin{subtable}[t]{\columnwidth}
	\centering
	\setlength\tabcolsep{4pt}
		\vspace{12pt}
	\caption{Dual-SANs converted voice} \label{table:1a2}
	\vspace{-3pt}
		\begin{tabular}{c|c c c c c c}
		\toprule
		\multicolumn{1}{c|}{Tar./Percept} & \multicolumn{1}{c}{Sad} & Angry & Surprise & High Tension & Happy &  Neutral \\
		\midrule
		Sad & \textbf{68} & 3 & 2 & 0  & 0 & 27 \\
		Angry & 2 & \textbf{61} & 3 & 15  & 0 & 19\\
		Surprise & 0 & 5 & \textbf{55} & 25  & 10 & 5 \\
		High tension & 0 & 3 & 10 & \textbf{45}  & 32 & 10 \\
		Happy & 0 & 0 & 7 & 32  & \textbf{48} & 13 \\
		\bottomrule
	\end{tabular}
	\end{subtable}

    \begin{subtable}[t]{\columnwidth}
	\centering
	\setlength\tabcolsep{4pt}
	\vspace{12pt}
	\caption{CycleGAN converted voice} \label{table:1a3}
	\vspace{-3pt}
		\begin{tabular}{c|c c c c c c}
\toprule
		\multicolumn{1}{c|}{Tar./Percept} & \multicolumn{1}{c}{Sad} & Angry & Surprise & High Tension & Happy &  Neutral \\
		\midrule
		Sad & \textbf{62} & 2 & 6 & 2  & 3 & 25 \\
		Angry & 5 & \textbf{40} & 8 & 22  & 0 & 25\\
		Surprise & 0 & 5 & \textbf{48} & 17  & 20 & 10 \\
		High tension & 0 & 10 & 15 & \textbf{44}  & 26 & 5 \\
		Happy & 0 & 2 & 10 & 26  & \textbf{42} & 20 \\
		\bottomrule
	\end{tabular}
	\end{subtable}

	 \begin{subtable}[t]{\columnwidth}
	\centering
	\setlength\tabcolsep{4pt}
	\vspace{12pt}
	\caption{SPEECHFLOW converted voice} \label{table:1a4}
	\vspace{-3pt}
		\begin{tabular}{c|c c c c c c}
		\toprule
		\multicolumn{1}{c|}{Tar./Percept} & \multicolumn{1}{c}{Sad} & Angry & Surprise & High Tension & Happy &  Neutral \\
		\midrule
		Sad & \textbf{66} & 0 & 0 & 0  & 0 & 1 \\
		Angry & 0 & \textbf{63} & 0 & 2  & 0 & 2\\
		Surprise & 0 & 0 & \textbf{54} & 2  & 2 & 2 \\
		High tension & 0 & 2 & 2 & \textbf{50}  & 6 & 0 \\
		Happy & 0 & 0 & 2 & 6  & \textbf{52} & 0 \\
		\bottomrule
	\end{tabular}
	\end{subtable}

	 \begin{subtable}[t]{\columnwidth}
	\centering
	\setlength\tabcolsep{4pt}
	\vspace{12pt}
	\caption{SFEVC converted voice} \label{table:1a5}
	\vspace{-3pt}
		\begin{tabular}{c|c c c c c c}
		\toprule
		\multicolumn{1}{c|}{Tar./Percept} & \multicolumn{1}{c}{Sad} & Angry & Surprise & High Tension & Happy &  Neutral \\
		\midrule
		Sad & \textbf{69} & 2 & 0 & 0  & 0 & 29 \\
		Angry & 5 & \textbf{66} & 2 & 22  & 0 & 5\\
		Surprise & 0 & 4 & \textbf{60} & 16  & 12 & 8 \\
		High tension & 0 & 7 & 7 & \textbf{55}  & 26 & 5 \\
		Happy & 0 & 3 & 12 & 22  & \textbf{58} & 5 \\
		\bottomrule
	\end{tabular}
	\end{subtable}

	 \begin{subtable}[t]{\columnwidth}
	\centering
	\setlength\tabcolsep{4pt}
	\vspace{12pt}
	\caption{SFEVC+VA converted voice} \label{table:1a6}
	\vspace{-3pt}
		\begin{tabular}{c|c c c c c c}
		\toprule
		\multicolumn{1}{c|}{Tar./Percept} & \multicolumn{1}{c}{Sad} & Angry & Surprise & High Tension & Happy &  Neutral \\
		\midrule
		Sad & \textbf{68} & 0 & 0 & 0  & 0 & 30 \\
		Angry & 3 & \textbf{69} & 3 & 18 & 0 & 7\\
		Surprise & 0 & 3 & \textbf{64} & 22  & 6 & 3 \\
		High tension & 0 & 8 & 3 & \textbf{60}  & 25 & 4 \\
		Happy & 0 & 2 & 6 & 26  & \textbf{62} & 4 \\
		\bottomrule
	\end{tabular}
	\end{subtable}
\end{table}

\subsection{Comparative Study}
To evaluate the proposed method, we reimplemented several state-of-the-art emotional VC models for comparison.
\begin{itemize}
    \item\textbf{DBNs+NNs:} This is the earliest emotional VC method based on deep learning models~\cite{luo2016emotional}. The model uses the DBNs to convert spectral features while using the NNs to convert the F0 features. 
    
     \item\textbf{Dual-SANs:} This model adopts the dual supervised  adversarial network, in which continuous wavelet transform method was used to augment prosody (F0) features.  
    
	\item\textbf{CycleGAN:} The CycleGAN model \cite{kaneko2017parallel} has been widely used in the non-parallel VC tasks. Kun \textit{et al.} \cite{zhou2020transforming} have also used this unsupervised learning model in the emotional VC. 
	
	\item\textbf{SPEECHFLOW:} This is a state-of-the-art source filter method that has been used in the voice conversion  \cite{qian2020unsupervised}. We applied it in the emotional VC that uses the designed bottlenecks auto-encoder for filtering the timbre and prosody features.
	
	\item\textbf{SFEVC:} This is our proposed method SFEVC that uses designed bottlenecks multi-channel encoders and the emotion-separate encoders, while without two-stage training.
	\item\textbf{SFEVC+VA:}
	We adopt the novel VA-based two-stages training to train our SFEVC, denoted as SFEVC+VA. This model was built to validate the effectiveness of the two-stage training.

\end{itemize}

\subsection{Objective Evaluation}
In the emotional VC task, the timbre and pitch features are mainly represented by mel-cepstral coefficients (MCC) and F0 features. Therefore, for objective evaluation, we use mel-cepstral distortion (MCD) to measure how close the converted MCC is to the target MCC in the mel-cepstral space, and root means square error (RMSE) to evaluate the conversion error between target F0 and converted F0. The MCD and F0 values are calculated using the different emotional voices from the same speaker since the emotional VC aims to convert the emotion-related features but keep the speaker identity unchanged. The averaged MCD and RMSE are shown in Table~\ref{tb:result2} and Table~\ref{tb:result3}, respectively. 


As shown in Table~\ref{tb:result2}, the MCD values between source and target emotion speech samples are over 3, indicating that the timbre features of different emotional voices spoken by the same speaker also have difference. Although, they are not as high as the values in the normal speaker VC task, which are close to 5 [], []. We also find that, comparing the MCD of the source voice and converted voice, all models show decreased values, but not significantly. Comparing Dual-SANs, SPEECHFLOW, SFEVC, and SFEVC+AV with CycleGAN, the MCD varies slightly for N2Sa but decreases significantly for the other high arousal emotions. This is because CycleGAN is effective for transferring sophisticated local texture appearance between image domains, but it has difficulties with objects that have both related appearance and shape changes. For the angry and happy voice conversion, their spectrum shape is more different from the neutral voice than the sad voice. This shows that although cycle-consistency is effective for the training of GANs when converting the neutral voice to a sad voice, the effect of cycle-consistency is no more than the dual supervised learning models (Dual-SANs) and the designed bottlenecks encoder models (SPEECHFLOW and SFEVC) in the emotional VC task, in which the emotional features' shapes change a lot.  



\begin{figure*}[h]
	\centering
	\includegraphics[width=0.9\linewidth]{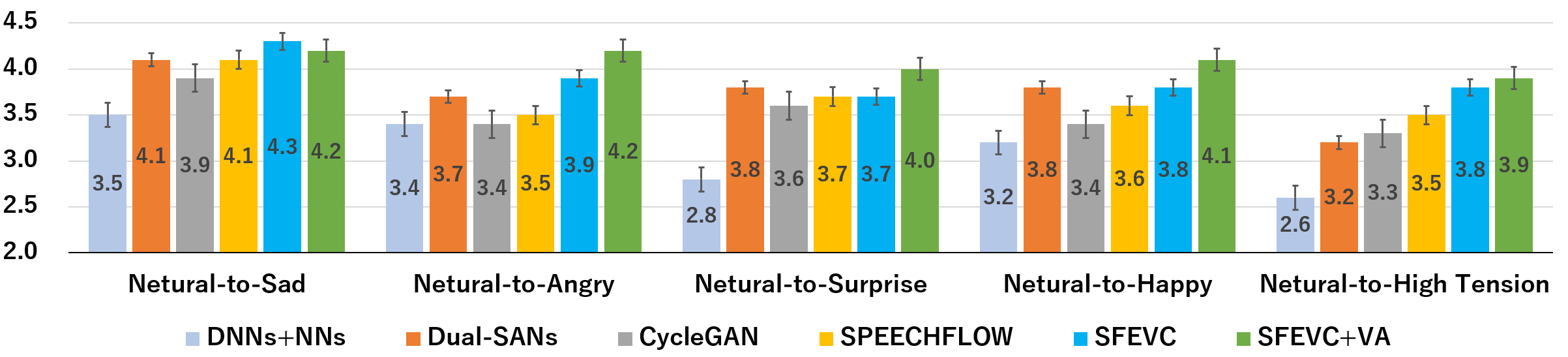}
	\centering
	\caption{MOS of the naturalness evaluation for the conversion of neutral voice to emotional voice, with 95\% confidence intervals computed from the t-test.}	
	\label{fig:NatureN2A}
	\vspace{-10pt}
\end{figure*}

As shown in Table~\ref{tb:result3}, comparing the RMSE results, all models show significantly decreased values. The bottlenecks encoder models can obtain better results than DBNs+NNs and the GAN-related models (Dual-SANs and CycleGANs). In the bottlenecks encoder models, our proposed SFEVC and SFEVC+VA models get better RMSE values than the SPEECHFLOW in the neutral to high tension emotion and surprise emotion. This indicates that our emotional separate encoder has efficiency in complex emotions. Comparing the results of SFEVC and SFEVC+VA, the SFEVC+VA has less error for the conversion from neutral to angry. 

\subsection{Subjective Evaluation}
For the subjective experiment, similarity test and naturalness MOS test are used as evaluation metrics.

\subsubsection{Similarity}
In the emotional VC task, the more similar the converted voice sounds to the target emotion, the more effective the model is. Therefore, we carry out a subjective emotion classification test for the neutral voice to emotion pairs including Neutral-to-Angry, Neutral-to-Sad, Neutral-to-Happy, Neutral-to-Surprise, and Neutral-to-High Tension, comparing different methods. 30 native-speaker listeners of different ages are involved and asked to label a converted voice as angry, sad, surprise, happy, high tension, or neutral emotion.

As shown in Table~\ref{tb:result5}(a), when evaluating the original recorded emotional speech utterances, all classification accuracy is higher than 90\%, indicating the classifier performs well enough to be used in the emotion classification test. The classification accuracy of Neutral-to-Sad is nearly 100\%, indicating that the human can easily separate the low valence emotion (sad) from the high valence voice (happy, angry, surprise, and high tension). Because the high tension and happy voice are similar, the classification accuracy of Neutral-to-Happy and Neutral-to-High Tension is lower than the others. The classification results for the converted voices of DBNs+NNs, Dual-SANs, CycleGAN, SPEECHFLOW, and SFEVC and SFEVC+VA are shown in (b), (c), (d), (e), (f), and (g) of Table~\ref{tb:result5}, respectively. 

As shown in Table~\ref{tb:result5} (b), the conventional DBNs+NNs method shows poor performance in all emotional VC tasks, especially for the conversion from neutral to angry and from neutral to high-tension. This result confirms that the DBNs+NNs model training without the GANs models (Dual-SANs and CycleGAN) or designed bottlenecks auto-encoder models (SPEECHFLOW, SFEVC, and SFEVC+VA) cannot convert the emotion features well.

Comparing the results of GANs based models (Dual-SANs and CycleGAN) with the source-filter models (SPEECHFLOW, SFEVC, and SFEVC+VA), the classification accuracy is similar for the converted angry voice and sad voice. However, get lower quality for the converted surprised voice and converted high-tension voice, indicating that the source-filter models get better conversion efficiency for the more complex emotions (surprise and high tension).

Comparing the results among the source-filter methods, i.e. Table~\ref{tb:result5} (e), (f) and (g), the proposed SFEVC has obtained better results than the SPEECHFLOW. Moreover, comparing the results of SFEVC and SFEVC+VA, it proved that applying the two stages training method based on 2D VA space can improve the quality for all the high valence emotion conversions.

\subsubsection{Naturalness}
In line with most previous works in the VC field, to measure naturalness, we conducted a MOS test for naturalness evaluation by 30 native subjects of different ages. The scale ranged from 1 (totally unnatural) to 5 (completely natural). The results are shown in Figure \ref{fig:NatureN2A}. In this test, a higher value indicates a better result, where the error bar shows the $95\%$ confidence interval. From these results, we can see that all naturalness scores are above or near 3, which means that reasonable naturalness. Comparing the results of different models, we can see that the GANs models and source-filter models improved a lot than the DNBs+NNs. In the GANs models, the Dual-SANs gets more stable results than the CycleGANs, which has low naturalness in the neutral to high-tension converted voice. The source-filter models can get better results than the GANs models, especially for the high arousal voice conversion. Our proposed SFEVC model obtains better results than the other source filter model. Especially for the conversion of neutral voice to high tension voice, the score is 0.3 higher than the other methods. 

\section{Conclusion}
\label{Sec:con}
In this work, we have presented a source-filter emotional voice conversion model applied with the multi-channel encoders and emotion-separate encoders, which can better decompose the timbre, pitch, content and rhythm features from the speech information and decouple the speaker-independent emotion features from the speaker-dependent emotion features. Moreover, we have also introduced a two-stage training strategy based on the valence-arousal space, which can improve the conversion expressiveness for the high arousal emotion. 
Experimental results show that our proposed SFEVC model is more effective in converting any emotions compared to the conventional method and achieved the state-of-the-art results.

\appendices

\section{DETAILED CALCULATION OF RECONSTRUCTING THE CONVERTED EMOTIONAL VOICE}

Assume that the speech $\boldsymbol{X}$ can be reconstructed by (C,R,P,T) as follows: 
\begin{equation}
    \label{equation:apendix}
     \boldsymbol{X}= D(C,R,P,T)                   =D(C,R,(P^D,P^I),(T^D,T^I))
\end{equation} 
where $C$, $R$, $P$, and $T$ represent the content, rhythm, pitch and timbre, respectively. The pitch and timbre features can be separated by speaker independent features ($P^I,T^I$) and speaker dependent features ($P^D,T^D$). In the emotional VC, for the input source emotion voice $\boldsymbol{X}_s= D(C_s,R_s,(P^D_s,P^I_s),(T^D_s,T^I_s))$, the main task is to replace speaker independent emotion features $(P^I_s,T^I_s)$ to the target emotion features $(P^I_t,T^I_t)$, but keep the other features unchanged as $\boldsymbol{X}_t=D(C_s,R_s,(P^D_s,P^I_t),(T^D_s,T^I_t))$. Combing Eq. 1, Eq. 2 and Eq. 3, our emotional VC task is to reconstruct the converted voice as follows:
\begin{equation}
\begin{aligned}
\hat{\boldsymbol{X}}_{s\rightarrow t}  
 = & D(\boldsymbol{Z}_{r},\boldsymbol{Z}_{c},\boldsymbol{Z}_{Z_{f}}^{t},\boldsymbol{Z}_{Z_{f}}^{s},\boldsymbol{Z}_{U}^{t},\boldsymbol{Z}_{U}^{s})\\
 = & D(E_{r}(\boldsymbol{X}),E_{c}(A(\boldsymbol{X})), E_{\boldsymbol{Z}_{f}}^{s}(\boldsymbol{Z}_{f}^{s},\boldsymbol{Z}_{f}^{y}),\\  & E_{\boldsymbol{Z}_{f}}^{t-s}   (\boldsymbol{Z}_{f}^{t},\boldsymbol{Z}_{f}^{s}), E_{\boldsymbol{U}}^{s}(\boldsymbol{U},\boldsymbol{U^y}), E_{\boldsymbol{U}}^{t-s}(\boldsymbol{U^{t},U}))\\
= &\boldsymbol{X}_t =  D(C_s,R_s,(P_s^D,P_t^I),(T_s^D,T_t^I)),
\end{aligned}
\end{equation}
which achieves 0 reconstruction loss in Eq. 4.

As the inputs are the same speech in different emotion ($\boldsymbol{X}_s$,$\boldsymbol{X}_t$) spoken by the same speaker X, the content ($C_s$), rhythm ($R_s$), and speaker dependent features ($P^D_s,T^D_s$) are the same information. Therefore, the loss function to minimize the $||\hat{\boldsymbol{X}}_{s\rightarrow t}-\boldsymbol{X}_t||$ depends on reconstructing the speaker independent emotion features ($P^I,T^I$) by encoders ($E_{\boldsymbol{Z}_{f}}^{t-s}$ $E_{\boldsymbol{U}}^{t-s}$) and decoder $D$ as follows:
\begin{equation}
\begin{split}
\min_{E_{\boldsymbol{Z}_{f}}^{t-s}(\cdot), E_{\boldsymbol{U}}^{t-s}(\cdot),D(\cdot )}\mathbb{L}||\hat{\boldsymbol{X}}_{s\rightarrow t}-\boldsymbol{X}_t||
= \mathbb{L}_{P^I} + \lambda_{1} \mathbb{L}_{T^I},
\end{split}
\end{equation}
where, 
\begin{equation}
\begin{split}
\mathbb{L}_{P^I}&=\mathbb{E} [|| E_{\boldsymbol{Z}_{f}}^{t-s}(\boldsymbol{Z}_{f}^{t},\boldsymbol{Z}_{f}^{s})-P_t^I ||_{1}]\\
 &= \mathbb{E} [||E_{\boldsymbol{Z}_{f}}^{t-s}(E_{f}(A(\boldsymbol{P})),E_{f}(A(\boldsymbol{P^{t}}))-P_t^I ||_{1}],\\
\mathbb{L}_{T^I} &= \mathbb{E} [|| E_{\boldsymbol{U}}^{t-s}(\boldsymbol{U^{t},U})-T_t^I ||_{1}]
\end{split}
\end{equation}
where, the weight $\lambda_{1}$ is set to 1.


As the inputs are the same speech in the same emotion ($\boldsymbol{X}_s,\boldsymbol{Y}_s$) by different speakers X and Y. The target speech can be represented as follows:
\begin{equation}
\begin{aligned}
\boldsymbol{X}_t=\boldsymbol{Y}_s=D(C_s,R_s,(P^D_y,P^I_s),(T^D_y,T^I_s)) 
\end{aligned}
\end{equation}
where, $C_s,R_s,P^I_s,T^I_s$ are the same information as $\boldsymbol{X}_s$. Our emotional VC task is to keep the speaker dependent features decoupled from timbre and pitch features of $\boldsymbol{X}_s$ unchanged. Then, the loss function to minimize the $||\hat{\boldsymbol{X}}_{s\rightarrow t}-\boldsymbol{X}_t||$ depends on reconstructing the speaker dependent features ($P^D,T^D$) as follows:

\begin{equation}
\begin{split}
\min_{E_{\boldsymbol{Z}_{f}}^{s}(\cdot ), E_{\boldsymbol{U}}^{s}(\cdot ),D(\cdot )}\mathbb{L}||\hat{\boldsymbol{X}}_{s\rightarrow t}-\boldsymbol{X}_t||
= \mathbb{L}_{P^D} +  \lambda_{2} \mathbb{L}_{T^D},
\end{split}
\end{equation}
where, 
\begin{equation}
\begin{split}
\mathbb{L}_{P^D}&=\mathbb{E} [|| E_{\boldsymbol{Z}_{f}}^{s}(\boldsymbol{Z}_{f}^{s},\boldsymbol{Z}_{f}^{y})-P_s^D ||_{1}]\\
 &= \mathbb{E} [||E_{\boldsymbol{Z}_{f}}^{s}(E_{f}(A(\boldsymbol{P})),E_{f}(A(\boldsymbol{P^{y}}))-P_s^D ||_{1}],\\
\mathbb{L}_{T^D} &= \mathbb{E} [|| E_{\boldsymbol{U}}^{s}(\boldsymbol{U,U^{y}})-T_s^D ||_{1}]
\end{split}
\end{equation}
where, the weight $\lambda_{2}$ is set to 1.



\bibliographystyle{IEEEtran}
\bibliography{mybib}




\end{document}